\documentclass[prl,twocolumn,showpacs,floatfix,aps]{revtex4-1}     
\usepackage{graphicx}           
\usepackage{amssymb}
\usepackage{natbib}
\usepackage{color}
\usepackage[bottom]{footmisc}
\begin{document}
\title{Entropy-driven  impurity-induced nematic-isotropic transition of liquid crystals}

\author{Pritam Kumar Jana$^{1,2}$}
\email{pritam.jana@aalto.fi}
\author{Julien Lam$^2$ }
%\altaffiliation{J.L. and N.P. contributed equally to this work.}
\author{Nagma Parveen$^3$} 
%\altaffiliation{J.L. and N.P. contributed equally to this work.}
\author{Mikko J Alava$^1$}
\author{Lasse Laurson$^{4}$}
\email{lasse.laurson@tuni.fi}

\affiliation{$^1$COMP Centre of Excellence, Department of Applied Physics, Aalto University, Espoo, Finland\\
$^2$Center for Nonlinear Phenomena and Complex Systems, Universite Libre de Bruxelles, 1050 Brussels, Belgium\\
$^3$Laboratory for Photochemistry and Spectroscopy, Department of Chemistry, KU Leuven, Leuven, Belgium\\
%$^4$State Key Laboratory of Multiphase Complex Systems, Chinese Academy of Sciences (CAS), Beijing PR China\\
$^4$Computational Physics Laboratory, Tampere University, P.O. Box 692, FI-33014 Tampere, Finland}

\begin{abstract}
Phase behavior of liquid crystals is of long-standing interest due to numerous applications,
with one of the key issues being how the presence of impurities affects the liquid crystalline order.
Here we study the orientational order of 4-cyano-4$^{'}$-pentylbiphenyl (5CB) and 4-cyano-4$^{'}$-hexylbiphenyl (6CB) nematic liquid crystals in the presence of 
varying concentrations of water and n-hexane molecules serving as impurities, by carrying out 
both fully atomistic simulations and experiments. 
Our results reveal that mixing of the impurities (in case of hexane) with the host liquid crystals 
causes a nematic-to-isotropic phase transition with hexane concentration as the control parameter while 
demixing (in case of water) results in only weak impurity-induced perturbations to the nematic 
liquid-crystalline order. We develop a coarse-grained model illustrating the general nature and entropic origin 
of the mixing-induced phase transition.
\end{abstract}

\maketitle

%\textit{Introduction}-\,
Liquid crystals (LCs) are recognized as a fascinating class of soft condensed matter 
because of their counter-intuitive behavior with fluidity and long 
range-ordering \cite{prost1995physics}.
LCs and their phase behavior are relevant both in many natural systems, as well as
for a wide range of technological applications.  
For instance, many features observed in living systems are reminiscent of liquid 
crystalline texture and to model them a description of liquid crystalline order 
is taken into account \cite{bouligand2008liquid,reyliquid}. 
From application perspective, while LCs were originally employed in flat panel 
displays \cite{stannarius2009liquid}, they are currently considered also in lots of 
other applications, including organic electronics \cite{nazarenko2010lyotropic}, optics, 
nano-/micro manipulation \cite{lagerwall2012new}, electronics, 
bio-sensing \cite{shiyanovskii2005lyotropic}, novel composites, and compact 
lenses \cite{liswitchable}. All of those applications rely on the orientationally 
ordered state of the LCs called the nematic phase \cite{martire1979nematic, shah2008field, 
chen2008isotropic, sidky2018silico, poulin1997novel}. 

While pure nematic LCs are widely studied, the properties of LC-based 
complex fluids, i.e., mixtures of LCs and other components, are less investigated 
even if related applications exist \cite{poulin1997novel}. In such systems new 
features can arise due to combining physical 
properties of each element or from new structural organizations \cite{prost1995physics}. 
Yamamoto {\it et al.} \cite{yamamototransparent} have studied mixtures of water, 
5CB LCs %in the nematic state 
and surfactants, and observed a novel phase, labeled the 'transparent nematic' phase, 
where on large scales the LC system appears isotropic while locally exhibiting
nematic order. Poulin {\it et al.} \cite{poulin1997novel} have observed an intriguing colloidal 
interaction, originating from the balance of dipole-dipole attractive and 
defect-mediated repulsive interactions which leads to the formation of chains 
of water droplets in nematic host-5CB LCs. 
Organization of anisotropic 1D nano-materials \cite{Hegmann2007,Bisoyi} or carbon 
nanotubes \cite{Dierking} dispersed in thermotropic LCs is the focus of current 
research because controlling the order in such systems leads to novel electro-optical 
applications. In these systems, nematic order of the LCs induces order of the 
anisotropic particles. However, phase behaviour of LCs can be modified when anisotropic molecules/colloids are dispersed in the LC medium.
To understand such phase behaviour, Katleen {\it {et al.}} \cite{denolfeffect} 
performed experiments with 4-n-octyl-4$^{'}$-cyanobiphenyl (8CB) LCs in the presence of cyclohexane and water and 
measured the nematic-to-isotropic phase transition temperature. A linear decrease 
of the transition temperature is observed as the mole fraction of cyclohexane is 
increased while water weakly affects the transition temperature. These results 
were explained using macroscopic arguments based on Landau-de Gennes theory along 
with an additional term coupling order parameter and mole fraction of the added 
impurities. Yet, at this stage, description of the relevant microscopic mechanisms 
are largely missing in the literature as it is difficult to observe experimentally 
such phase transitions at the atomistic or molecular level. 

\begin{figure*}[]
    \centering
        \centering
        \includegraphics[scale=0.55]{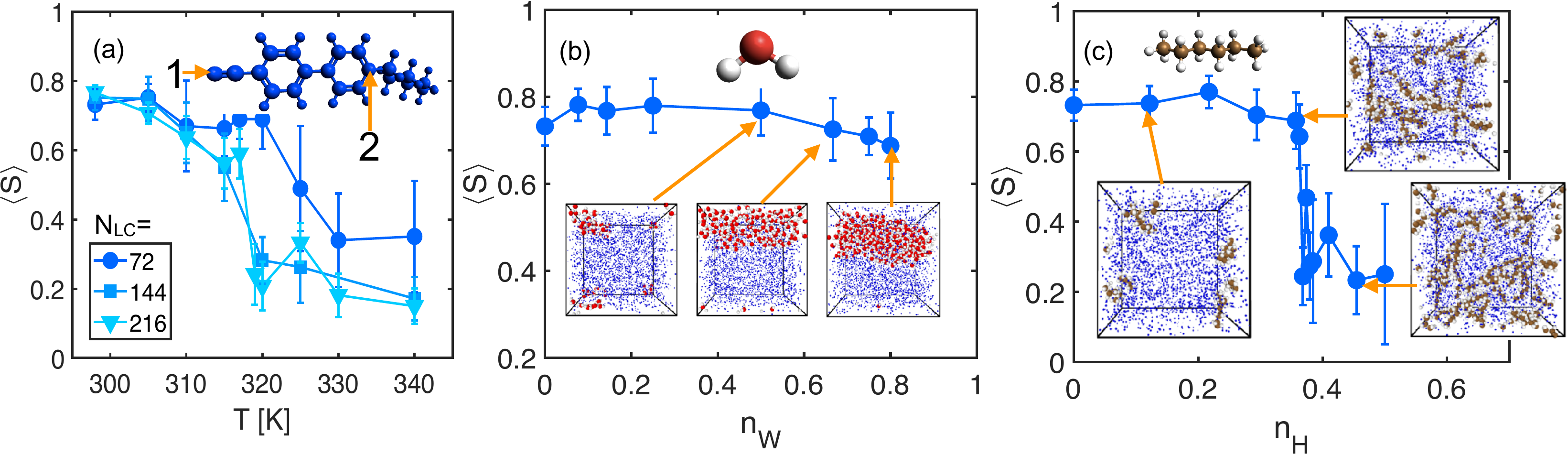} 
   \caption{Results from all-atom simulations. (a) Order parameter $\langle S \rangle$ as a 
function of temperature $T$ for three different system sizes with $N_\textrm{LC} = 72, 144, 
\text{and } 216$. The transition temperature found in the simulations is slightly larger than the experimental 
value $T_\textrm{NI} = 303.3$ K \cite{kobinata,vieweg2010molecular} for 6CB, possibly due to
finite size effects. The inset shows the structure of 6CB LC molecule, indicating the two 
atoms (1 and 2) which define the long axis of the molecule. 
(b) $\langle S \rangle$ as a function of water concentration at  
$T = 298$ K. The insets display snapshots of the LC-water mixtures at water concentrations 
$\eta_\textrm{W} = 0.5, 0.67, \text{and } 0.8$ (from left to right). (c) $\langle S \rangle$ as a 
function of hexane concentration at $T = 298$ K. 
The insets display snapshots of the LC-hexane mixtures at hexane concentrations
$\eta_\textrm{H} = 0.12, 0.36, \text{and } 0.45$. In (b) and (c) the number of LCs 
$N_\textrm{LC}$ is kept fixed at 72. The blue points in the snapshots of (b) and (c) 
represent all atoms of the LCs molecules. Lines are guide to the eyes.}
    \label{Fig:Fig1}
\end{figure*}

In this article, we combine
fully atomistic simulations and experiments to show how 
nematic orientational order of the LCs is influenced by the 
presence of various concentrations of impurities of different kinds.  
We start by performing atomistic simulations of a 
typical LC, 4-cyano-4$^{'}$-hexylbiphenyl (6CB). Two types of impurities 
mixed with the 6CB LC in its nematic phase are considered: water, 
which is an associative liquid, and hexane, which is non-associative. 
Our results show that water tends to form droplets within the LC, 
and hence does not significantly affect the nematic order of the LC.
In contrast to this, adding hexane will lead to strong mixing of 
hexane and the LC, resulting in an impurity-induced 
nematic-to-isotropic transition at a critical hexane concentration. 
These numerical predictions are then confirmed 
by experiments with a similar system: 5 CB LCs mixed with 
different concentrations of water or hexane. Finally, a 
coarse-grained model is developed to extract the main physical 
features responsible for our results, illustrating the entropic 
origin of the impurity-induced nematic-to-isotropic transition. 
Our main result is twofold: (i) Ordering of LCs is only weakly
perturbed by impurities which phase separate into droplets, while 
impurities mixing with the LCs may induce a nematic-isotropic 
phase transion, and (ii) such mixing and demixing within the LC 
host is driven simply by the presence or absence of attractive 
interactions between the impurity molecules.

Several efforts have been made to simulate properties of liquid 
crystal systems such as 4-cyano-4-alkylbiphenyl (nCB). In 
particular, the so-called odd-even effect in nematic-isotropic 
transition was recently reproduced up to a satisfactory level of 
accuracy in computer simulations \cite{tiberio2009towards}. 
In the present study we simulate 6CB, but homologues should lead 
to qualitatively similar results. 
To construct the atomistic models of 6CB and hexane molecules, we 
use force field parameters from Refs. \cite{adam,cheung,janananoscale}. For 
water, the SPC/E model is employed. Details of the force fields 
are given in the Supplemental Material. With this 
model, molecular dynamics simulations are performed with a time 
step of 1\,fs. At the initial state, the system is randomly 
disposed in a large simulation box of linear dimensions 
$L_x=L_y=L_z$ = 70 \AA and equilibrated at a high temperature 
of $T=360$ K for $10$ ps under NVT conditions. Then, the 
temperature is reduced with a constant rate to the target 
temperature $T_\textrm{f}$. Calculations are finally performed 
for another $40$\,ns under NPT conditions at 1 atm and results 
are computed from time-averaging over the last $5$\,ns.

\begin{figure*}[]
    \centering
        \centering
        \includegraphics[scale=0.55]{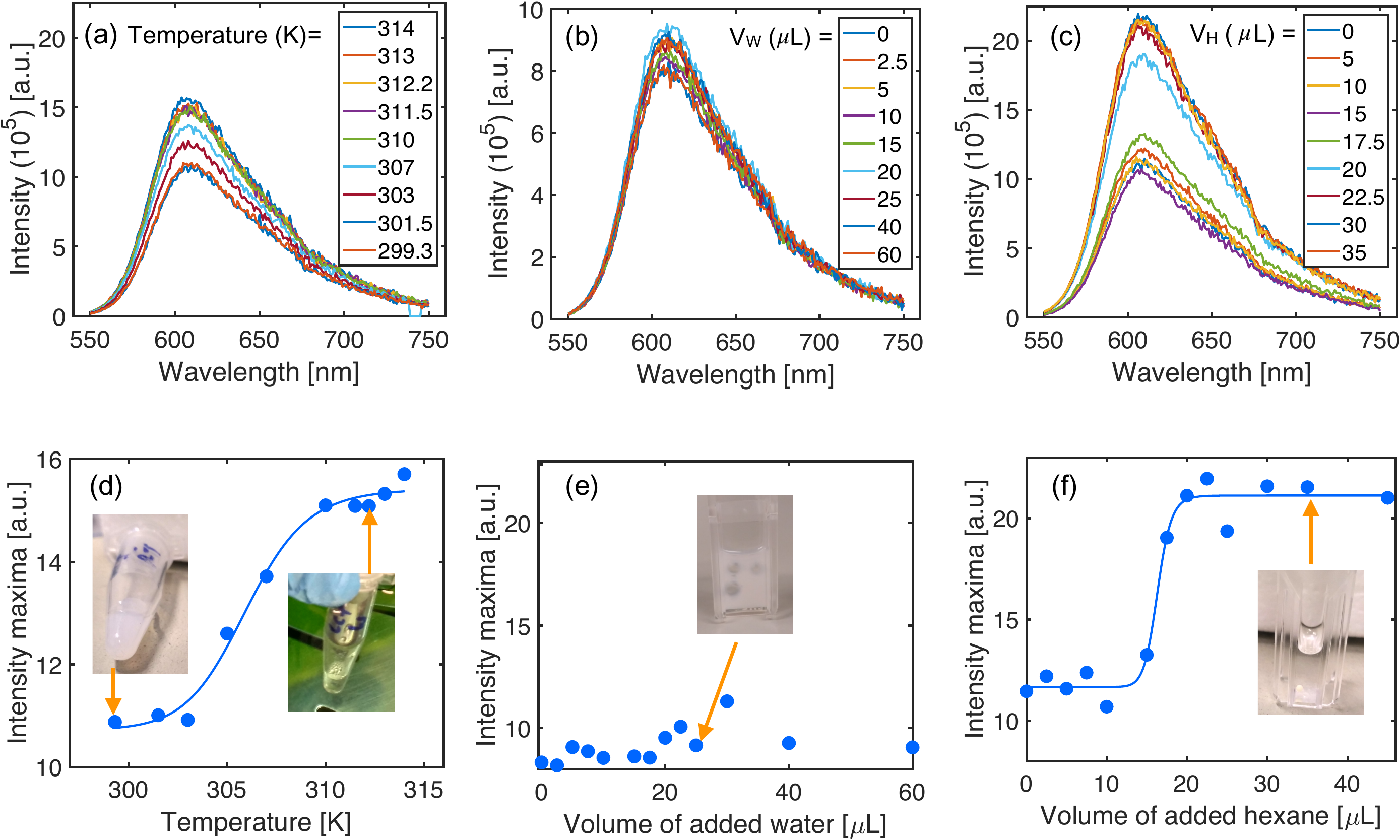} 
   \caption{Results from experiments. Fluorescence emission spectra 
of Nile blue (excitation at 525 nm non-polarized light) in 5CB LC 
acquired (a) at different temperatures (b) upon addition of water 
and (c) upon addition of hexane at the specified volumes. Peak maxima 
of the corresponding spectra as a function of (d) temperature, 
(e) added volume of water and (f) added volume of hexane. The images 
in the inset display different physical states (turbid [left side of 
(d)], transparent [right side of (d) and (f)] and immiscible [(e)]) 
of the 5CB LC. In (d) and (f), lines are guide to the eyes.}
\label{Fig:Fig2}
\end{figure*}

To characterize the resulting LC configurations, the order parameter $S$ of the LCs is 
computed as the maximum eigenvalue of the average ordering tensor $Q_{\alpha,\beta}$,
\begin{equation}
Q_{\alpha,\beta}=\frac{1}{N}\sum_i^N\left(\frac{3}{2}u_\alpha^i u_\beta^i-\frac{1}{2}\delta_{\alpha,\beta}\right),
\end{equation}
where $u_\alpha^i ({\alpha,\beta}=x,y,z)$ are the Cartesian components of the unit 
vector of the LC molecule $i$, $N$ is the number of LC molecules, and 
$\delta_{\alpha,\beta}$ is the Kronecker delta. Here, the unit vector of the LC 
molecule is taken to be parallel to the line connecting the atom 1 and atom 2 as 
shown in the inset of Fig.\ref{Fig:Fig1}(a). With this formulation, having a large value 
of $S$ means that most of the LCs are oriented in the same direction. 

Before studying the influence of impurities, the temperature-dependent orientational 
order of pure 6CB LC is characterized in Fig.\ref{Fig:Fig1}(a). In particular, a typical 
nematic to isotropic transition is observed as the temperature increases starting 
from the low-temperature nematic phase. The largest studied system 
contains $N_\textrm{LC}=216$\,LC molecules and exhibits a transition at 
$T \approx 315$\,K. In what follows, the influence of impurities is studied at room 
temperature ($T=298$\,K) where the pure system remains nematic. In general, one 
expects a $2D$ 'phase diagram' $S(T,n)$, with $n$ the impurity concentration.

When adding increasing numbers $N_\textrm{W}$ of water molecules to the system of 
$N_\textrm{LC}=72$ LC molecules, the LCs have a strong tendency to demix with water 
resulting in formation of water droplets of different sizes at different water 
concentrations $n_\textrm{W} \equiv N_\textrm{W}/(N_\textrm{W}+N_\textrm{LC})$; 
large, system-spanning droplets at large water concentrations appear as 'tubes' due 
periodic boundaries, see the insets of  Fig.\ref{Fig:Fig1}(b).
It is well-established that water molecules form intermolecular hydrogen bonds of 
energy 23.4 KJ mol$^{-1}$ \cite{suresh2000hydrogen} between electronegative oxygen and 
electropositive hydrogen atoms via which a network stretching throughout the liquid is 
constructed and a droplet is formed \cite{errington2001relationship}. 
Fig.\ref{Fig:Fig1}(b) shows that the presence of such water droplet(s) only 
slightly perturbs 
the ordering of the LCs. On the other hand, when adding hexane to the initially nematic LC, 
the scenario is completely different. Indeed, hexane and liquid crystals strongly mix  
due to hexane being a non-polar molecule because of 2 factors: (i) the only C-H 
bond present in hexane is non-polar due to the very similar electronegativities, and (ii) 
hexane is symmetric such that any polarity in the molecule would cancel out. Therefore, 
the only intermolecular force acting in hexane are induced dipole-dipole forces or van 
der Waals forces/London dispersion forces which are weak and short-ranged, preventing the 
aggregation of hexane in the LC medium, 
see the inset of Fig.\ref{Fig:Fig1}(c). This mixing of the hexane impurities with the LC 
molecules results in a sharp nematic-to-isotropic transition at a critical hexane 
concentration of $n_\textrm{H} \equiv N_\textrm{H}/(N_\textrm{H}+N_\textrm{LC})\approx 0.4$, 
see Fig.\ref{Fig:Fig1}(c). 

In order to confirm these numerical predictions of how ordering of LC 
molecules is affected by the presence of water and hexane, experiments are 
carried out. 4-cyano-4-pentylbiphenyl (5CB) instead of 6CB was studied to 
emphasize the universality of the observed mechanisms. At 299\,K, the pure 
system exhibits clear nematic ordering which is characterized by a turbid white 
color. Upon heating, the system becomes transparent, implying a transition from 
the nematic to the isotropic phase, see the insets of 
Fig.\ref{Fig:Fig2}(d). In order to 
quantitatively probe this transition, nile blue molecules are added to 
the system as their fluorescence properties are sensitive to the LC 
micro-environment \cite{tajalliphotophysical}. Indeed, the nile blue emission 
intensity is higher in the isotropic phase because of either an increased 
fluorescence absorption or a change in the dipole-moment orientation in 
the optically excited state.

\begin{figure*}[]
    \centering
        \centering
        \includegraphics[scale=0.55]{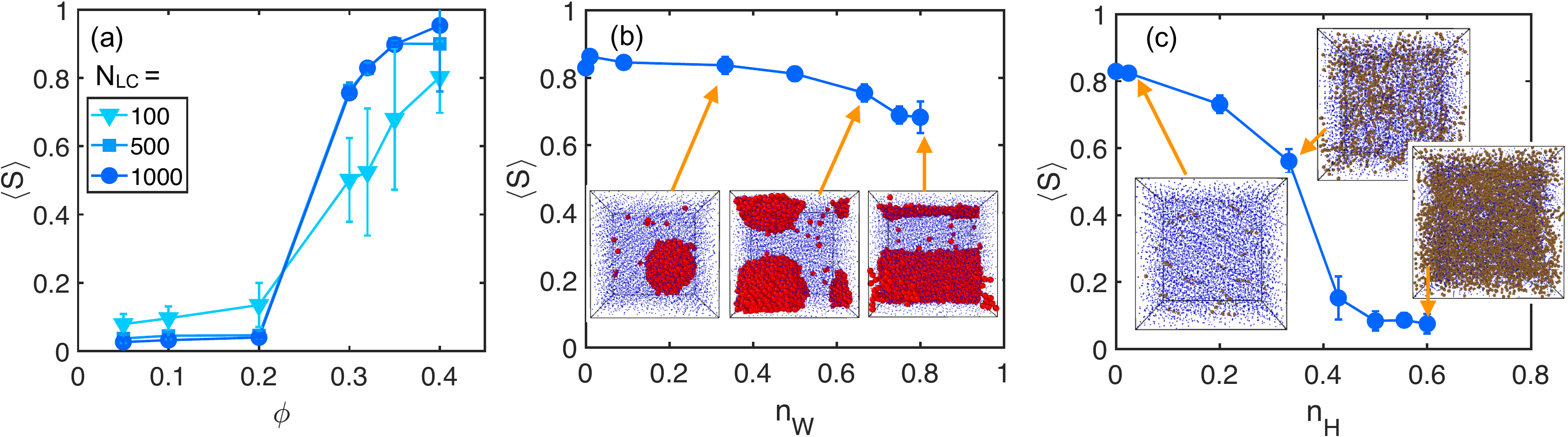}
   \caption{Results from coarse-grained simulations. (a) $\langle S \rangle$ as a
function of packing fraction $\phi$ for coarse-grained simulations for three
different system sizes with $N_\textrm{LC} = 100, 500$, and 1000. (b) $\langle S \rangle$ 
as a function of water concentration at $\phi=0.32$. The insets display
snapshots of the LC-water mixtures at water concentrations $\eta_W = 0.5, 0.65, 
\text{and } 0.8$. (c) $\langle S \rangle$ as a function of hexane concentration
at $\phi=0.32$. The insets display snapshots of the LC-hexane mixtures at hexane 
concentrations $\eta_H = 0.02, 0.33, \text{and } 0.6$. In (b) and (c), the number 
of LCs $N_{LC}$ is kept fixed at 1000. The blue points in the snapshots of (b) and 
(c) represent the beads of LCs. Lines are guide to the eyes.}
\label{Fig:Fig3}
\end{figure*}

First, in the pure LC solution, the fluorescence emission signal increases upon 
heating as seen in Fig.\ref{Fig:Fig2}(a). Note that upon cooling the 5CBs from 315 K 
to 299 K, the reverse behavior 
is also observed. In Fig.\ref{Fig:Fig2}(d), the nematic to isotropic phase transition of 5CB is 
observed at approximately 307\,K (comparable to the previous experiment \cite{shimfluorescence}),
which is close to our atomistic simulation results and 
therefore confirms the accuracy of our modeling. Upon addition of water to the LC system we 
again observe a good agreement with the atomistic simulations as no significant evolution 
of the fluorescence emission with water concentration is spotted [Fig.\ref{Fig:Fig2}(c)]. 
Furthermore, water droplets can also be seen in the experiments [inset of Fig.\ref{Fig:Fig2}(e)]. 
Finally, as predicted by our atomistic simulations, an impurity-induced nematic-to-isotropic 
transition is observed when adding hexane molecules at a fixed temperature 
of 299\,K(Fig.\ref{Fig:Fig2}(b)). 
The transition is sharp, and occurs at about 20 $\mu$l of added hexane in 325 $\mu$l of 5CB, 
which translates into $n_H$ of 0.12 (Fig.\ref{Fig:Fig2}(e)). In comparison to 
the atomistic simulations with 6CB, the transition occurs at a lower hexane concentration. 
Such a difference is expected to result from the small difference between the molecular 
chain length of 5CB and 6CB, differences in the reduced temperature 
$(T_\textrm{c}-T)/T_\textrm{c}$ in the simulations and experiment, as well as from 
some inherent impurities of the commercial 5CB studied (98\% pure). Overall, the key 
features predicted by our atomistic simulations are well-reproduced in experiments. 

To better understand the difference between water and hexane in LCs observed both 
in atomistic simulations and experiments, we develop and study a coarse-grained model 
of our LC-impurity system \cite{PhysRevE.90.012404,tian2001molecular}.
The model ignores many molecular details, focusing 
on the role of energy and entropy in the phase behaviour, and allows to avoid 
finite-size effects as we can easily simulate 
much larger systems of up to 1000 LC molecules, see Fig. SM3 in the Supplemental Material. In the model, LC molecules are modeled as 8 purely repulsive 
hard-sphere beads. The number of beads is adjusted in order to match the size ratio 
between 6CB and water molecules. Each bead of a LC molecule interacts with the beads 
of the other molecules using a cut-and-shifted Lennard-Jones (LJ) potential including only the 
repulsive part. Intra-chain beads interact via a harmonic potential with a high spring 
constant of $2000 \epsilon/\sigma^2$ for both stretching and angle coefficients such that 
the chains behave like rigid rods. The hexane molecules are modeled like the LCs but 
with 3 beads instead of 8. 
Water molecules are considered as attractive particles using 
the same cut-and-shifted LJ potential as above but with a larger cutoff. 
As such, by defining $\sigma$ as the length parameter of the LJ potential and 
thus the size of a water molecule, the only remaining parameter is the employed cutoff which 
is equal to 1.122$\sigma$ for the repulsive interaction (hard sphere) and 2.5$\sigma$ for 
the attractive interaction (water). For the purely repulsive particles, temperature scale 
is not relevant. Therefore, NVE simulations are used in order to prevent any issues due to 
adding a thermostat. Attractive water molecules are 
simulated using the NVT ensemble with $k_\textrm{B}T=0.8\epsilon$ in order to reasonably describe 
water at 300K \cite{Lam2}, with $k_\textrm{B}$ the Boltzmann constant. At the initialization 
step, particles are randomly disposed inside a large simulation box that is reduced using 
a fixed compression rate in order to match the target packing fraction 
$\phi = (1/V)\cdot N_\mathrm{LC}\cdot n_\mathrm{bead}\cdot (4/3)\pi R_\mathrm{LC}^3$
(with $V$ is the volume of the box, $R_\mathrm{LC}$ is the radius and $n_\mathrm{bead}$ the 
number of beads within each LC which is equal to 8)
which plays the role of the inverse temperature in our hard-sphere model (notice that increasing 
$\phi$ and inverse temperature both lead to a decrease of entropy)\,\cite{PhysRevLett}; 
LCs interacting via hard interactions with $\phi$ as the control parameter
have been studied by several groups to explore the phase
behaviour of LCs \cite{williamson1998liquid,vroegephase,onsager1949effects}. 
Then, simulations are continued for $10^7$ steps to reach equilibrium.
We point out that while in attractive systems ordering transitions can be obtained by 
both concentration increase or a decrease in temperature, in purely repulsive systems 
they can be observed only by using concentration as the control parameter. 

This model is able to reproduce the isotropic-nematic transition of the pure LCs  by increasing $\phi$ 
instead of reducing temperature [Fig. \ref{Fig:Fig3}(a)]. 
Notice that analogously to what was found in our atomistic simulations, there is a shift 
and rounding of the the transition (a finite size effect) when the number of molecules is 
reduced. From here, we choose to work at $\phi=0.32$ which is large enough to be located 
in the nematic phase without suffering from equilibration issues that emerge at larger 
packing fractions. The agreement with atomistic simulations and experiments continues in 
the presence of coarse-grained water molecules (Fig. \ref{Fig:Fig3}(b)). This can be
expected as similarly to the full atomistic simulations, also here formation of water 
droplets with the LC is observed, originating from the attractive part in the water-water 
interaction potential [compare Fig. \ref{Fig:Fig3}(b) with Fig. \ref{Fig:Fig1}(b)]. 
Finally, adding coarse-grained hexane randomly dispersed within the LC
medium perturbs the LC order parameter similarly to our observations in the atomistic 
simulations and experiments, resulting in an impurity-induced nematic-isotropic transition 
at $n_\textrm{H} \approx 0.4$ [Fig. \ref{Fig:Fig3} (c)]. 

In conclusion, we studied the influence of adding impurities to a nematic LC system 
using both atomistic simulations and experiments. Depending on the nature of the 
impurities, two distinct behaviors are observed. For water, droplet formation 
(phase separation) takes place resulting in only weak perturbations to the nematic
LC order, while for hexane, no demixing is observed and a sharp decrease of the nematic order parameter is found for large 
enough hexane concentrations. These results suggest a more general conclusion: 
adding non-mesogenic impurities should induce a nematic-to-isotropic phase transition 
only when the impurities mix with the host LCs. This picture is 
supported by the results from our coarse-grained model, where the 
structural complexity of the molecules is reduced by modelling them as rigid hard 
chain-type (LC and hexane) or single-bead (water) molecules. The model thus highlights 
the crucial role of the presence or absense of mixing of the impurities with the LCs 
in determining if the impurities are able to significantly perturb the LC order, 
and that in the case of the purely repulsive coarse-grained LC-hexane model, the 
impurity-induced nematic-isotropic transition is mostly driven by entropic effects.
Hence, this work not only contributes to the overall understanding of the role of 
impurities mixed with LCs, but also provides a benchmark for controlling LC phase 
behavior by the addition of non-interacting impurities. 

We acknowledge the support of the Academy of Finland through the Centres of 
Excellence Programme (2012-2017, project no. 251748; PKJ, MJA and LL), the FiDiPro 
Programme (project no. 13282993, PKJ), and an Academy Research Fellowship (project 
no. 268302, LL). 
 JL acknowledges financial support of the Fonds de la
Recherche Scientifique - FNRS. We acknowledge the computational resources provided 
by the Aalto University School of Science ``Science-IT'' project, CSC (Finland), 
Consortium des Equipements de Calcul Intensif (CECI), and the F\'ed\'eration 
Lyonnaise de Mod\'elisation et Sciences Num\'eriques (FLMSN). 

J.L. and N.P. contributed equally to this work.
%\end{acknowledgments}
%\vspace{2 cm}
%J.L. and N.P. contributed equally to this work.
%\bibliographystyle{unsrt} 
%\bibliography{biblio.bib}

\begin{thebibliography}{35}%
\makeatletter
\providecommand \@ifxundefined [1]{%
 \@ifx{#1\undefined}
}%
\providecommand \@ifnum [1]{%
 \ifnum #1\expandafter \@firstoftwo
 \else \expandafter \@secondoftwo
 \fi
}%
\providecommand \@ifx [1]{%
 \ifx #1\expandafter \@firstoftwo
 \else \expandafter \@secondoftwo
 \fi
}%
\providecommand \natexlab [1]{#1}%
\providecommand \enquote  [1]{``#1''}%
\providecommand \bibnamefont  [1]{#1}%
\providecommand \bibfnamefont [1]{#1}%
\providecommand \citenamefont [1]{#1}%
\providecommand \href@noop [0]{\@secondoftwo}%
\providecommand \href [0]{\begingroup \@sanitize@url \@href}%
\providecommand \@href[1]{\@@startlink{#1}\@@href}%
\providecommand \@@href[1]{\endgroup#1\@@endlink}%
\providecommand \@sanitize@url [0]{\catcode `\\12\catcode `\$12\catcode
  `\&12\catcode `\#12\catcode `\^12\catcode `\_12\catcode `\%12\relax}%
\providecommand \@@startlink[1]{}%
\providecommand \@@endlink[0]{}%
\providecommand \url  [0]{\begingroup\@sanitize@url \@url }%
\providecommand \@url [1]{\endgroup\@href {#1}{\urlprefix }}%
\providecommand \urlprefix  [0]{URL }%
\providecommand \Eprint [0]{\href }%
\providecommand \doibase [0]{http://dx.doi.org/}%
\providecommand \selectlanguage [0]{\@gobble}%
\providecommand \bibinfo  [0]{\@secondoftwo}%
\providecommand \bibfield  [0]{\@secondoftwo}%
\providecommand \translation [1]{[#1]}%
\providecommand \BibitemOpen [0]{}%
\providecommand \bibitemStop [0]{}%
\providecommand \bibitemNoStop [0]{.\EOS\space}%
\providecommand \EOS [0]{\spacefactor3000\relax}%
\providecommand \BibitemShut  [1]{\csname bibitem#1\endcsname}%
\let\auto@bib@innerbib\@empty
%</preamble>
\bibitem [{\citenamefont {Prost}(1995)}]{prost1995physics}%
  \BibitemOpen
  \bibfield  {author} {\bibinfo {author} {\bibfnamefont {J.}~\bibnamefont
  {Prost}},\ }\href@noop {} {\emph {\bibinfo {title} {The Physics of Liquid
  Crystals}}},\ Vol.~\bibinfo {volume} {83}\ (\bibinfo  {publisher} {Oxford
  university press},\ \bibinfo {year} {1995})\BibitemShut {NoStop}%
\bibitem [{\citenamefont {Bouligand}(2008)}]{bouligand2008liquid}%
  \BibitemOpen
  \bibfield  {author} {\bibinfo {author} {\bibfnamefont {Y.}~\bibnamefont
  {Bouligand}},\ }\href@noop {} {\bibfield  {journal} {\bibinfo  {journal}
  {Comptes Rendus Chimie}\ }\textbf {\bibinfo {volume} {11}},\ \bibinfo {pages}
  {281} (\bibinfo {year} {2008})}\BibitemShut {NoStop}%
\bibitem [{\citenamefont {Rey}(2010)}]{reyliquid}%
  \BibitemOpen
  \bibfield  {author} {\bibinfo {author} {\bibfnamefont {A.~D.}\ \bibnamefont
  {Rey}},\ }\href@noop {} {\bibfield  {journal} {\bibinfo  {journal} {Soft
  Matter}\ }\textbf {\bibinfo {volume} {6}},\ \bibinfo {pages} {3402} (\bibinfo
  {year} {2010})}\BibitemShut {NoStop}%
\bibitem [{\citenamefont {Stannarius}(2009)}]{stannarius2009liquid}%
  \BibitemOpen
  \bibfield  {author} {\bibinfo {author} {\bibfnamefont {R.}~\bibnamefont
  {Stannarius}},\ }\href@noop {} {\bibfield  {journal} {\bibinfo  {journal}
  {Nat. Mater.}\ }\textbf {\bibinfo {volume} {8}},\ \bibinfo {pages} {617}
  (\bibinfo {year} {2009})}\BibitemShut {NoStop}%
\bibitem [{\citenamefont {Nazarenko}\ \emph {et~al.}(2010)\citenamefont
  {Nazarenko}, \citenamefont {Boiko}, \citenamefont {Anisimov}, \citenamefont
  {Kadashchuk}, \citenamefont {Nastishin}, \citenamefont {Golovin},\ and\
  \citenamefont {Lavrentovich}}]{nazarenko2010lyotropic}%
  \BibitemOpen
  \bibfield  {author} {\bibinfo {author} {\bibfnamefont {V.}~\bibnamefont
  {Nazarenko}}, \bibinfo {author} {\bibfnamefont {O.}~\bibnamefont {Boiko}},
  \bibinfo {author} {\bibfnamefont {M.}~\bibnamefont {Anisimov}}, \bibinfo
  {author} {\bibfnamefont {A.}~\bibnamefont {Kadashchuk}}, \bibinfo {author}
  {\bibfnamefont {Y.~A.}\ \bibnamefont {Nastishin}}, \bibinfo {author}
  {\bibfnamefont {A.}~\bibnamefont {Golovin}}, \ and\ \bibinfo {author}
  {\bibfnamefont {O.}~\bibnamefont {Lavrentovich}},\ }\href@noop {} {\bibfield
  {journal} {\bibinfo  {journal} {Appl. Phys. Lett.}\ }\textbf {\bibinfo
  {volume} {97}},\ \bibinfo {pages} {284} (\bibinfo {year} {2010})}\BibitemShut
  {NoStop}%
\bibitem [{\citenamefont {Lagerwall}\ and\ \citenamefont
  {Scalia}(2012)}]{lagerwall2012new}%
  \BibitemOpen
  \bibfield  {author} {\bibinfo {author} {\bibfnamefont {J.~P.}\ \bibnamefont
  {Lagerwall}}\ and\ \bibinfo {author} {\bibfnamefont {G.}~\bibnamefont
  {Scalia}},\ }\href@noop {} {\bibfield  {journal} {\bibinfo  {journal} {Curr.
  Appl. Phys.}\ }\textbf {\bibinfo {volume} {12}},\ \bibinfo {pages} {1387}
  (\bibinfo {year} {2012})}\BibitemShut {NoStop}%
\bibitem [{\citenamefont {Shiyanovskii}\ \emph {et~al.}(2005)\citenamefont
  {Shiyanovskii}, \citenamefont {Lavrentovich}, \citenamefont {Schneider},
  \citenamefont {Ishikawa}, \citenamefont {Smalyukh}, \citenamefont
  {Woolverton}, \citenamefont {Niehaus},\ and\ \citenamefont
  {Doane}}]{shiyanovskii2005lyotropic}%
  \BibitemOpen
  \bibfield  {author} {\bibinfo {author} {\bibfnamefont {S.}~\bibnamefont
  {Shiyanovskii}}, \bibinfo {author} {\bibfnamefont {O.}~\bibnamefont
  {Lavrentovich}}, \bibinfo {author} {\bibfnamefont {T.}~\bibnamefont
  {Schneider}}, \bibinfo {author} {\bibfnamefont {T.}~\bibnamefont {Ishikawa}},
  \bibinfo {author} {\bibfnamefont {I.}~\bibnamefont {Smalyukh}}, \bibinfo
  {author} {\bibfnamefont {C.}~\bibnamefont {Woolverton}}, \bibinfo {author}
  {\bibfnamefont {G.}~\bibnamefont {Niehaus}}, \ and\ \bibinfo {author}
  {\bibfnamefont {K.}~\bibnamefont {Doane}},\ }\href@noop {} {\bibfield
  {journal} {\bibinfo  {journal} {Mol. Cryst. Liq. Cryst.}\ }\textbf {\bibinfo
  {volume} {434}},\ \bibinfo {pages} {259} (\bibinfo {year}
  {2005})}\BibitemShut {NoStop}%
\bibitem [{\citenamefont {Li}\ \emph {et~al.}(2006)\citenamefont {Li},
  \citenamefont {Mathine}, \citenamefont {Valley}, \citenamefont
  {{\"A}yr{\"a}s}, \citenamefont {Haddock}, \citenamefont {Giridhar},
  \citenamefont {Williby}, \citenamefont {Schwiegerling}, \citenamefont
  {Meredith}, \citenamefont {Kippelen}, \citenamefont {Honkanen},\ and\
  \citenamefont {Peyghambarian}}]{liswitchable}%
  \BibitemOpen
  \bibfield  {author} {\bibinfo {author} {\bibfnamefont {G.}~\bibnamefont
  {Li}}, \bibinfo {author} {\bibfnamefont {D.~L.}\ \bibnamefont {Mathine}},
  \bibinfo {author} {\bibfnamefont {P.}~\bibnamefont {Valley}}, \bibinfo
  {author} {\bibfnamefont {P.}~\bibnamefont {{\"A}yr{\"a}s}}, \bibinfo {author}
  {\bibfnamefont {J.~N.}\ \bibnamefont {Haddock}}, \bibinfo {author}
  {\bibfnamefont {M.}~\bibnamefont {Giridhar}}, \bibinfo {author}
  {\bibfnamefont {G.}~\bibnamefont {Williby}}, \bibinfo {author} {\bibfnamefont
  {J.}~\bibnamefont {Schwiegerling}}, \bibinfo {author} {\bibfnamefont {G.~R.}\
  \bibnamefont {Meredith}}, \bibinfo {author} {\bibfnamefont {B.}~\bibnamefont
  {Kippelen}}, \bibinfo {author} {\bibfnamefont {S.}~\bibnamefont {Honkanen}},
  \ and\ \bibinfo {author} {\bibfnamefont {N.}~\bibnamefont {Peyghambarian}},\
  }\href@noop {} {\bibfield  {journal} {\bibinfo  {journal} {Proc. Natl. Acad.
  Sci. U.S.A.}\ }\textbf {\bibinfo {volume} {103}},\ \bibinfo {pages} {6100}
  (\bibinfo {year} {2006})}\BibitemShut {NoStop}%
\bibitem [{\citenamefont {Martire}\ and\ \citenamefont
  {Dowell}(1979)}]{martire1979nematic}%
  \BibitemOpen
  \bibfield  {author} {\bibinfo {author} {\bibfnamefont {D.}~\bibnamefont
  {Martire}}\ and\ \bibinfo {author} {\bibfnamefont {F.}~\bibnamefont
  {Dowell}},\ }\href@noop {} {\bibfield  {journal} {\bibinfo  {journal} {J.
  Chem. Phys.}\ }\textbf {\bibinfo {volume} {70}},\ \bibinfo {pages} {5914}
  (\bibinfo {year} {1979})}\BibitemShut {NoStop}%
\bibitem [{\citenamefont {Shah}\ \emph {et~al.}(2008)\citenamefont {Shah},
  \citenamefont {Fontecchio}, \citenamefont {Mattia},\ and\ \citenamefont
  {Gogotsi}}]{shah2008field}%
  \BibitemOpen
  \bibfield  {author} {\bibinfo {author} {\bibfnamefont {H.~J.}\ \bibnamefont
  {Shah}}, \bibinfo {author} {\bibfnamefont {A.~K.}\ \bibnamefont
  {Fontecchio}}, \bibinfo {author} {\bibfnamefont {D.}~\bibnamefont {Mattia}},
  \ and\ \bibinfo {author} {\bibfnamefont {Y.}~\bibnamefont {Gogotsi}},\
  }\href@noop {} {\bibfield  {journal} {\bibinfo  {journal} {J. Appl. Phys.}\
  }\textbf {\bibinfo {volume} {103}},\ \bibinfo {pages} {064314} (\bibinfo
  {year} {2008})}\BibitemShut {NoStop}%
\bibitem [{\citenamefont {Chen}\ \emph {et~al.}(2008)\citenamefont {Chen},
  \citenamefont {Hamlington},\ and\ \citenamefont {Shen}}]{chen2008isotropic}%
  \BibitemOpen
  \bibfield  {author} {\bibinfo {author} {\bibfnamefont {X.}~\bibnamefont
  {Chen}}, \bibinfo {author} {\bibfnamefont {B.~D.}\ \bibnamefont
  {Hamlington}}, \ and\ \bibinfo {author} {\bibfnamefont {A.~Q.}\ \bibnamefont
  {Shen}},\ }\href@noop {} {\bibfield  {journal} {\bibinfo  {journal}
  {Langmuir}\ }\textbf {\bibinfo {volume} {24}},\ \bibinfo {pages} {541}
  (\bibinfo {year} {2008})}\BibitemShut {NoStop}%
\bibitem [{\citenamefont {Sidky}\ \emph {et~al.}(2018)\citenamefont {Sidky},
  \citenamefont {de~Pablo},\ and\ \citenamefont {Whitmer}}]{sidky2018silico}%
  \BibitemOpen
  \bibfield  {author} {\bibinfo {author} {\bibfnamefont {H.}~\bibnamefont
  {Sidky}}, \bibinfo {author} {\bibfnamefont {J.~J.}\ \bibnamefont {de~Pablo}},
  \ and\ \bibinfo {author} {\bibfnamefont {J.~K.}\ \bibnamefont {Whitmer}},\
  }\href@noop {} {\bibfield  {journal} {\bibinfo  {journal} {Phys. Rev. Lett.}\
  }\textbf {\bibinfo {volume} {120}},\ \bibinfo {pages} {107801} (\bibinfo
  {year} {2018})}\BibitemShut {NoStop}%
\bibitem [{\citenamefont {Poulin}\ \emph {et~al.}(1997)\citenamefont {Poulin},
  \citenamefont {Stark}, \citenamefont {Lubensky},\ and\ \citenamefont
  {Weitz}}]{poulin1997novel}%
  \BibitemOpen
  \bibfield  {author} {\bibinfo {author} {\bibfnamefont {P.}~\bibnamefont
  {Poulin}}, \bibinfo {author} {\bibfnamefont {H.}~\bibnamefont {Stark}},
  \bibinfo {author} {\bibfnamefont {T.}~\bibnamefont {Lubensky}}, \ and\
  \bibinfo {author} {\bibfnamefont {D.}~\bibnamefont {Weitz}},\ }\href@noop {}
  {\bibfield  {journal} {\bibinfo  {journal} {Science}\ }\textbf {\bibinfo
  {volume} {275}},\ \bibinfo {pages} {1770} (\bibinfo {year}
  {1997})}\BibitemShut {NoStop}%
\bibitem [{\citenamefont {Yamamoto}\ and\ \citenamefont
  {Tanaka}(2001)}]{yamamototransparent}%
  \BibitemOpen
  \bibfield  {author} {\bibinfo {author} {\bibfnamefont {J.}~\bibnamefont
  {Yamamoto}}\ and\ \bibinfo {author} {\bibfnamefont {H.}~\bibnamefont
  {Tanaka}},\ }\href@noop {} {\bibfield  {journal} {\bibinfo  {journal}
  {Nature}\ }\textbf {\bibinfo {volume} {409}},\ \bibinfo {pages} {321}
  (\bibinfo {year} {2001})}\BibitemShut {NoStop}%
\bibitem [{\citenamefont {Hegmann}\ \emph {et~al.}(2007)\citenamefont
  {Hegmann}, \citenamefont {Qi},\ and\ \citenamefont {Marx}}]{Hegmann2007}%
  \BibitemOpen
  \bibfield  {author} {\bibinfo {author} {\bibfnamefont {T.}~\bibnamefont
  {Hegmann}}, \bibinfo {author} {\bibfnamefont {H.}~\bibnamefont {Qi}}, \ and\
  \bibinfo {author} {\bibfnamefont {V.~M.}\ \bibnamefont {Marx}},\ }\href@noop
  {} {\bibfield  {journal} {\bibinfo  {journal} {J. Inorg. Organomet. Polym.
  Mater.}\ }\textbf {\bibinfo {volume} {17}},\ \bibinfo {pages} {483} (\bibinfo
  {year} {2007})}\BibitemShut {NoStop}%
\bibitem [{\citenamefont {Bisoyi}\ and\ \citenamefont {Kumar}(2011)}]{Bisoyi}%
  \BibitemOpen
  \bibfield  {author} {\bibinfo {author} {\bibfnamefont {H.~K.}\ \bibnamefont
  {Bisoyi}}\ and\ \bibinfo {author} {\bibfnamefont {S.}~\bibnamefont {Kumar}},\
  }\href@noop {} {\bibfield  {journal} {\bibinfo  {journal} {Chem. Soc. Rev.}\
  }\textbf {\bibinfo {volume} {40}},\ \bibinfo {pages} {306} (\bibinfo {year}
  {2011})}\BibitemShut {NoStop}%
\bibitem [{\citenamefont {Dierking}\ \emph {et~al.}(2004)\citenamefont
  {Dierking}, \citenamefont {Scalia}, \citenamefont {Morales},\ and\
  \citenamefont {LeClere}}]{Dierking}%
  \BibitemOpen
  \bibfield  {author} {\bibinfo {author} {\bibfnamefont {I.}~\bibnamefont
  {Dierking}}, \bibinfo {author} {\bibfnamefont {G.}~\bibnamefont {Scalia}},
  \bibinfo {author} {\bibfnamefont {P.}~\bibnamefont {Morales}}, \ and\
  \bibinfo {author} {\bibfnamefont {D.}~\bibnamefont {LeClere}},\ }\href@noop
  {} {\bibfield  {journal} {\bibinfo  {journal} {Adv. Mater.}\ }\textbf
  {\bibinfo {volume} {16}},\ \bibinfo {pages} {865} (\bibinfo {year}
  {2004})}\BibitemShut {NoStop}%
\bibitem [{\citenamefont {Denolf}\ \emph {et~al.}(2007)\citenamefont {Denolf},
  \citenamefont {Cordoyiannis}, \citenamefont {Glorieux},\ and\ \citenamefont
  {Thoen}}]{denolfeffect}%
  \BibitemOpen
  \bibfield  {author} {\bibinfo {author} {\bibfnamefont {K.}~\bibnamefont
  {Denolf}}, \bibinfo {author} {\bibfnamefont {G.}~\bibnamefont
  {Cordoyiannis}}, \bibinfo {author} {\bibfnamefont {C.}~\bibnamefont
  {Glorieux}}, \ and\ \bibinfo {author} {\bibfnamefont {J.}~\bibnamefont
  {Thoen}},\ }\href@noop {} {\bibfield  {journal} {\bibinfo  {journal} {Phys.
  Rev. E}\ }\textbf {\bibinfo {volume} {76}},\ \bibinfo {pages} {051702}
  (\bibinfo {year} {2007})}\BibitemShut {NoStop}%
\bibitem [{\citenamefont {Kobinata}\ \emph {et~al.}(1986)\citenamefont
  {Kobinata}, \citenamefont {Kobayashi}, \citenamefont {Yoshida}, \citenamefont
  {Chandani},\ and\ \citenamefont {Maeda}}]{kobinata}%
  \BibitemOpen
  \bibfield  {author} {\bibinfo {author} {\bibfnamefont {S.}~\bibnamefont
  {Kobinata}}, \bibinfo {author} {\bibfnamefont {T.}~\bibnamefont {Kobayashi}},
  \bibinfo {author} {\bibfnamefont {H.}~\bibnamefont {Yoshida}}, \bibinfo
  {author} {\bibfnamefont {A.}~\bibnamefont {Chandani}}, \ and\ \bibinfo
  {author} {\bibfnamefont {S.}~\bibnamefont {Maeda}},\ }\href@noop {}
  {\bibfield  {journal} {\bibinfo  {journal} {J. Mol. Struct.}\ }\textbf
  {\bibinfo {volume} {146}},\ \bibinfo {pages} {373} (\bibinfo {year}
  {1986})}\BibitemShut {NoStop}%
\bibitem [{\citenamefont {Vieweg}\ \emph {et~al.}(2010)\citenamefont {Vieweg},
  \citenamefont {Jansen}, \citenamefont {Shakfa}, \citenamefont {Scheller},
  \citenamefont {Krumbholz}, \citenamefont {Wilk}, \citenamefont {Mikulics},\
  and\ \citenamefont {Koch}}]{vieweg2010molecular}%
  \BibitemOpen
  \bibfield  {author} {\bibinfo {author} {\bibfnamefont {N.}~\bibnamefont
  {Vieweg}}, \bibinfo {author} {\bibfnamefont {C.}~\bibnamefont {Jansen}},
  \bibinfo {author} {\bibfnamefont {M.~K.}\ \bibnamefont {Shakfa}}, \bibinfo
  {author} {\bibfnamefont {M.}~\bibnamefont {Scheller}}, \bibinfo {author}
  {\bibfnamefont {N.}~\bibnamefont {Krumbholz}}, \bibinfo {author}
  {\bibfnamefont {R.}~\bibnamefont {Wilk}}, \bibinfo {author} {\bibfnamefont
  {M.}~\bibnamefont {Mikulics}}, \ and\ \bibinfo {author} {\bibfnamefont
  {M.}~\bibnamefont {Koch}},\ }\href@noop {} {\bibfield  {journal} {\bibinfo
  {journal} {Opt. Express}\ }\textbf {\bibinfo {volume} {18}},\ \bibinfo
  {pages} {6097} (\bibinfo {year} {2010})}\BibitemShut {NoStop}%
\bibitem [{\citenamefont {Tiberio}\ \emph {et~al.}(2009)\citenamefont
  {Tiberio}, \citenamefont {Muccioli}, \citenamefont {Berardi},\ and\
  \citenamefont {Zannoni}}]{tiberio2009towards}%
  \BibitemOpen
  \bibfield  {author} {\bibinfo {author} {\bibfnamefont {G.}~\bibnamefont
  {Tiberio}}, \bibinfo {author} {\bibfnamefont {L.}~\bibnamefont {Muccioli}},
  \bibinfo {author} {\bibfnamefont {R.}~\bibnamefont {Berardi}}, \ and\
  \bibinfo {author} {\bibfnamefont {C.}~\bibnamefont {Zannoni}},\ }\href@noop
  {} {\bibfield  {journal} {\bibinfo  {journal} {ChemPhysChem}\ }\textbf
  {\bibinfo {volume} {10}},\ \bibinfo {pages} {125} (\bibinfo {year}
  {2009})}\BibitemShut {NoStop}%
\bibitem [{\citenamefont {Adam}\ \emph {et~al.}(1997)\citenamefont {Adam},
  \citenamefont {Clark}, \citenamefont {Ackland},\ and\ \citenamefont
  {Crain}}]{adam}%
  \BibitemOpen
  \bibfield  {author} {\bibinfo {author} {\bibfnamefont {C.}~\bibnamefont
  {Adam}}, \bibinfo {author} {\bibfnamefont {S.}~\bibnamefont {Clark}},
  \bibinfo {author} {\bibfnamefont {G.}~\bibnamefont {Ackland}}, \ and\
  \bibinfo {author} {\bibfnamefont {J.}~\bibnamefont {Crain}},\ }\href@noop {}
  {\bibfield  {journal} {\bibinfo  {journal} {Phys. Rev. E}\ }\textbf {\bibinfo
  {volume} {55}},\ \bibinfo {pages} {5641} (\bibinfo {year}
  {1997})}\BibitemShut {NoStop}%
\bibitem [{\citenamefont {Cheung}\ \emph {et~al.}(2002)\citenamefont {Cheung},
  \citenamefont {Clark},\ and\ \citenamefont {Wilson}}]{cheung}%
  \BibitemOpen
  \bibfield  {author} {\bibinfo {author} {\bibfnamefont {D.}~\bibnamefont
  {Cheung}}, \bibinfo {author} {\bibfnamefont {S.}~\bibnamefont {Clark}}, \
  and\ \bibinfo {author} {\bibfnamefont {M.~R.}\ \bibnamefont {Wilson}},\
  }\href@noop {} {\bibfield  {journal} {\bibinfo  {journal} {Phys. Rev. E}\
  }\textbf {\bibinfo {volume} {65}},\ \bibinfo {pages} {051709} (\bibinfo
  {year} {2002})}\BibitemShut {NoStop}%
\bibitem [{\citenamefont {Jana}\ \emph {et~al.}(2018)\citenamefont {Jana},
  \citenamefont {Chen}, \citenamefont {Alava},\ and\ \citenamefont
  {Laurson}}]{janananoscale}%
  \BibitemOpen
  \bibfield  {author} {\bibinfo {author} {\bibfnamefont {P.~K.}\ \bibnamefont
  {Jana}}, \bibinfo {author} {\bibfnamefont {W.}~\bibnamefont {Chen}}, \bibinfo
  {author} {\bibfnamefont {M.~J.}\ \bibnamefont {Alava}}, \ and\ \bibinfo
  {author} {\bibfnamefont {L.}~\bibnamefont {Laurson}},\ }\href@noop {}
  {\bibfield  {journal} {\bibinfo  {journal} {Phys. Chem. Chem. Phys.}\
  }\textbf {\bibinfo {volume} {20}},\ \bibinfo {pages} {18737} (\bibinfo {year}
  {2018})}\BibitemShut {NoStop}%
\bibitem [{\citenamefont {Suresh}\ and\ \citenamefont
  {Naik}(2000)}]{suresh2000hydrogen}%
  \BibitemOpen
  \bibfield  {author} {\bibinfo {author} {\bibfnamefont {S.}~\bibnamefont
  {Suresh}}\ and\ \bibinfo {author} {\bibfnamefont {V.}~\bibnamefont {Naik}},\
  }\href@noop {} {\bibfield  {journal} {\bibinfo  {journal} {J. Chem. Phys.}\
  }\textbf {\bibinfo {volume} {113}},\ \bibinfo {pages} {9727} (\bibinfo {year}
  {2000})}\BibitemShut {NoStop}%
\bibitem [{\citenamefont {Errington}\ and\ \citenamefont
  {Debenedetti}(2001)}]{errington2001relationship}%
  \BibitemOpen
  \bibfield  {author} {\bibinfo {author} {\bibfnamefont {J.~R.}\ \bibnamefont
  {Errington}}\ and\ \bibinfo {author} {\bibfnamefont {P.~G.}\ \bibnamefont
  {Debenedetti}},\ }\href@noop {} {\bibfield  {journal} {\bibinfo  {journal}
  {Nature}\ }\textbf {\bibinfo {volume} {409}},\ \bibinfo {pages} {318}
  (\bibinfo {year} {2001})}\BibitemShut {NoStop}%
\bibitem [{\citenamefont {Tajalli}\ \emph {et~al.}(2008)\citenamefont
  {Tajalli}, \citenamefont {Gilani}, \citenamefont {Zakerhamidi},\ and\
  \citenamefont {Tajalli}}]{tajalliphotophysical}%
  \BibitemOpen
  \bibfield  {author} {\bibinfo {author} {\bibfnamefont {H.}~\bibnamefont
  {Tajalli}}, \bibinfo {author} {\bibfnamefont {A.~G.}\ \bibnamefont {Gilani}},
  \bibinfo {author} {\bibfnamefont {M.}~\bibnamefont {Zakerhamidi}}, \ and\
  \bibinfo {author} {\bibfnamefont {P.}~\bibnamefont {Tajalli}},\ }\href@noop
  {} {\bibfield  {journal} {\bibinfo  {journal} {Dyes and Pigments}\ }\textbf
  {\bibinfo {volume} {78}},\ \bibinfo {pages} {15} (\bibinfo {year}
  {2008})}\BibitemShut {NoStop}%
\bibitem [{\citenamefont {Shim}\ \emph {et~al.}(2011)\citenamefont {Shim},
  \citenamefont {Kim}, \citenamefont {Kim},\ and\ \citenamefont
  {Oh-e}}]{shimfluorescence}%
  \BibitemOpen
  \bibfield  {author} {\bibinfo {author} {\bibfnamefont {T.}~\bibnamefont
  {Shim}}, \bibinfo {author} {\bibfnamefont {S.}~\bibnamefont {Kim}}, \bibinfo
  {author} {\bibfnamefont {D.}~\bibnamefont {Kim}}, \ and\ \bibinfo {author}
  {\bibfnamefont {M.}~\bibnamefont {Oh-e}},\ }\href@noop {} {\bibfield
  {journal} {\bibinfo  {journal} {J. Appl. Phys.}\ }\textbf {\bibinfo {volume}
  {110}},\ \bibinfo {pages} {063532} (\bibinfo {year} {2011})}\BibitemShut
  {NoStop}%
\bibitem [{\citenamefont {Chen}\ \emph {et~al.}(2014)\citenamefont {Chen},
  \citenamefont {Kulju}, \citenamefont {Foster}, \citenamefont {Alava},\ and\
  \citenamefont {Laurson}}]{PhysRevE.90.012404}%
  \BibitemOpen
  \bibfield  {author} {\bibinfo {author} {\bibfnamefont {W.}~\bibnamefont
  {Chen}}, \bibinfo {author} {\bibfnamefont {S.}~\bibnamefont {Kulju}},
  \bibinfo {author} {\bibfnamefont {A.~S.}\ \bibnamefont {Foster}}, \bibinfo
  {author} {\bibfnamefont {M.~J.}\ \bibnamefont {Alava}}, \ and\ \bibinfo
  {author} {\bibfnamefont {L.}~\bibnamefont {Laurson}},\ }\href {\doibase
  10.1103/PhysRevE.90.012404} {\bibfield  {journal} {\bibinfo  {journal} {Phys.
  Rev. E}\ }\textbf {\bibinfo {volume} {90}},\ \bibinfo {pages} {012404}
  (\bibinfo {year} {2014})}\BibitemShut {NoStop}%
\bibitem [{\citenamefont {Tian}\ \emph {et~al.}(2001)\citenamefont {Tian},
  \citenamefont {Bedrov}, \citenamefont {Smith},\ and\ \citenamefont
  {Glaser}}]{tian2001molecular}%
  \BibitemOpen
  \bibfield  {author} {\bibinfo {author} {\bibfnamefont {P.}~\bibnamefont
  {Tian}}, \bibinfo {author} {\bibfnamefont {D.}~\bibnamefont {Bedrov}},
  \bibinfo {author} {\bibfnamefont {G.~D.}\ \bibnamefont {Smith}}, \ and\
  \bibinfo {author} {\bibfnamefont {M.}~\bibnamefont {Glaser}},\ }\href@noop {}
  {\bibfield  {journal} {\bibinfo  {journal} {J. Chem. Phys.}\ }\textbf
  {\bibinfo {volume} {115}},\ \bibinfo {pages} {9055} (\bibinfo {year}
  {2001})}\BibitemShut {NoStop}%
\bibitem [{\citenamefont {Lam}\ and\ \citenamefont {Lutsko}(2018)}]{Lam2}%
  \BibitemOpen
  \bibfield  {author} {\bibinfo {author} {\bibfnamefont {J.}~\bibnamefont
  {Lam}}\ and\ \bibinfo {author} {\bibfnamefont {J.~F.}\ \bibnamefont
  {Lutsko}},\ }\href@noop {} {\bibfield  {journal} {\bibinfo  {journal} {J.
  Chem. Phys.}\ }\textbf {\bibinfo {volume} {149}},\ \bibinfo {pages} {134703}
  (\bibinfo {year} {2018})}\BibitemShut {NoStop}%
\bibitem [{\citenamefont {Sanz}\ \emph {et~al.}(2011)\citenamefont {Sanz},
  \citenamefont {Valeriani}, \citenamefont {Zaccarelli}, \citenamefont {Poon},
  \citenamefont {Pusey},\ and\ \citenamefont {Cates}}]{PhysRevLett}%
  \BibitemOpen
  \bibfield  {author} {\bibinfo {author} {\bibfnamefont {E.}~\bibnamefont
  {Sanz}}, \bibinfo {author} {\bibfnamefont {C.}~\bibnamefont {Valeriani}},
  \bibinfo {author} {\bibfnamefont {E.}~\bibnamefont {Zaccarelli}}, \bibinfo
  {author} {\bibfnamefont {W.~C.~K.}\ \bibnamefont {Poon}}, \bibinfo {author}
  {\bibfnamefont {P.~N.}\ \bibnamefont {Pusey}}, \ and\ \bibinfo {author}
  {\bibfnamefont {M.~E.}\ \bibnamefont {Cates}},\ }\href {\doibase
  10.1103/PhysRevLett.106.215701} {\bibfield  {journal} {\bibinfo  {journal}
  {Phys. Rev. Lett.}\ }\textbf {\bibinfo {volume} {106}},\ \bibinfo {pages}
  {215701} (\bibinfo {year} {2011})}\BibitemShut {NoStop}%
\bibitem [{\citenamefont {Williamson}\ and\ \citenamefont
  {Jackson}(1998)}]{williamson1998liquid}%
  \BibitemOpen
  \bibfield  {author} {\bibinfo {author} {\bibfnamefont {D.~C.}\ \bibnamefont
  {Williamson}}\ and\ \bibinfo {author} {\bibfnamefont {G.}~\bibnamefont
  {Jackson}},\ }\href@noop {} {\bibfield  {journal} {\bibinfo  {journal} {J.
  Chem. Phys.}\ }\textbf {\bibinfo {volume} {108}},\ \bibinfo {pages} {10294}
  (\bibinfo {year} {1998})}\BibitemShut {NoStop}%
\bibitem [{\citenamefont {Vroege}\ and\ \citenamefont
  {Lekkerkerker}(1992)}]{vroegephase}%
  \BibitemOpen
  \bibfield  {author} {\bibinfo {author} {\bibfnamefont {G.~J.}\ \bibnamefont
  {Vroege}}\ and\ \bibinfo {author} {\bibfnamefont {H.~N.}\ \bibnamefont
  {Lekkerkerker}},\ }\href@noop {} {\bibfield  {journal} {\bibinfo  {journal}
  {Rep. Prog. Phys.}\ }\textbf {\bibinfo {volume} {55}},\ \bibinfo {pages}
  {1241} (\bibinfo {year} {1992})}\BibitemShut {NoStop}%
\bibitem [{\citenamefont {Onsager}(1949)}]{onsager1949effects}%
  \BibitemOpen
  \bibfield  {author} {\bibinfo {author} {\bibfnamefont {L.}~\bibnamefont
  {Onsager}},\ }\href@noop {} {\bibfield  {journal} {\bibinfo  {journal} {Ann.
  N. Y. Acad. Sci.}\ }\textbf {\bibinfo {volume} {51}},\ \bibinfo {pages} {627}
  (\bibinfo {year} {1949})}\BibitemShut {NoStop}%
\end{thebibliography}

%\includegraphics{./Supplemental_Material.pdf}

\begin{figure}[h]
    \centering
    \includegraphics[width=20 cm, height=28 cm]{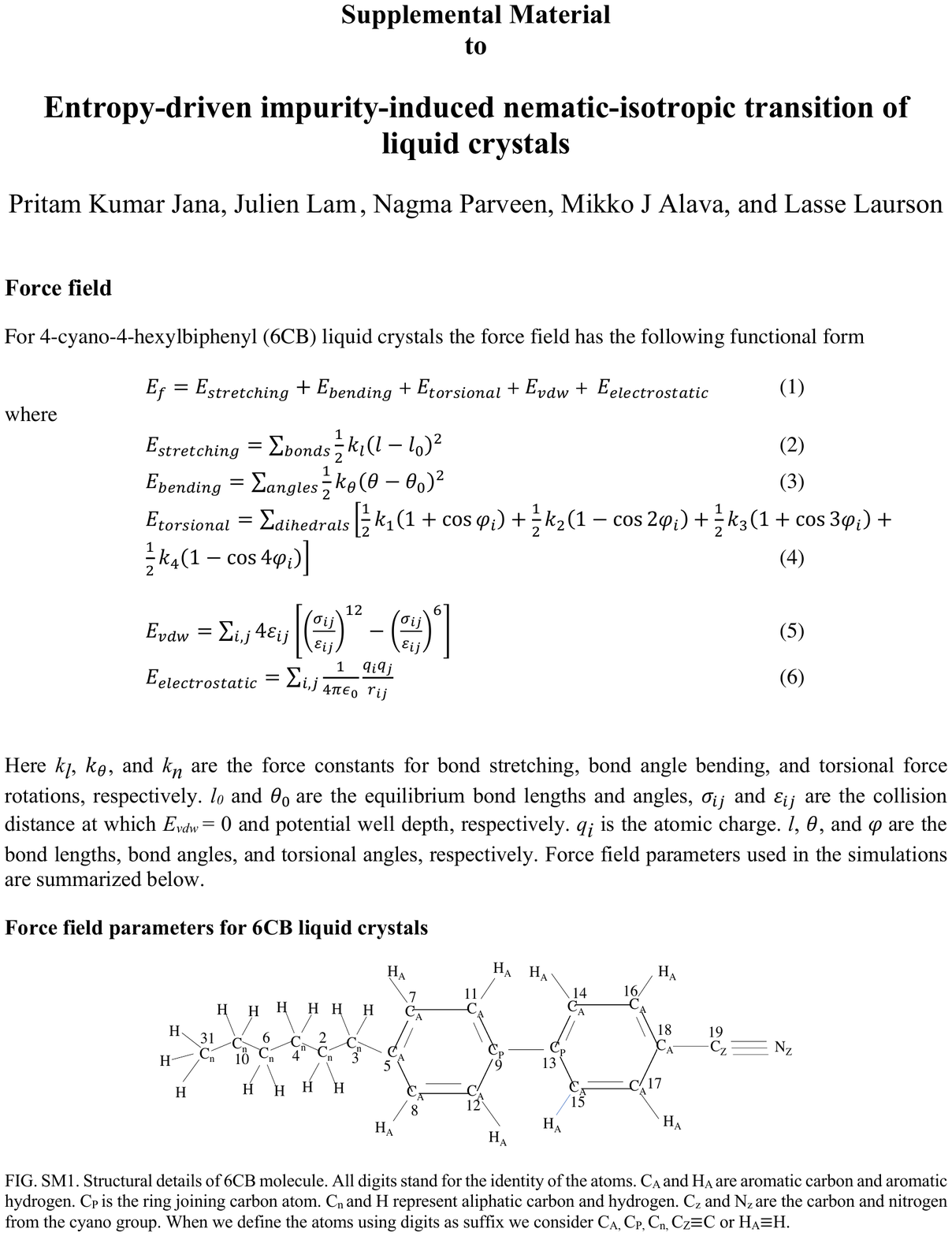}
\end{figure}
\begin{figure}[h]
    \centering
    \includegraphics[width=20 cm, height=28 cm]{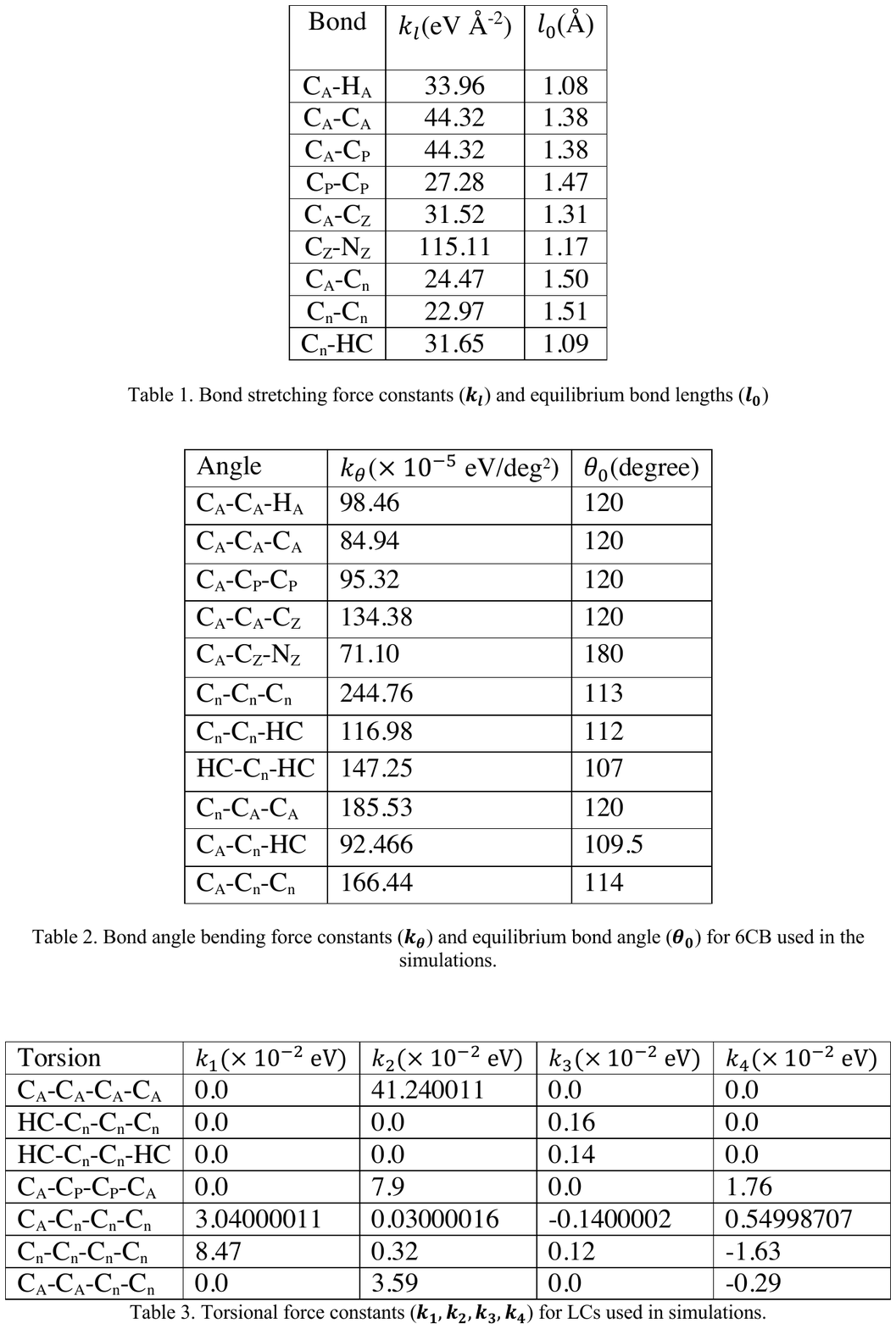} 
\end{figure}
\begin{figure}[h]
    \centering
    \includegraphics[width=20 cm, height=28 cm]{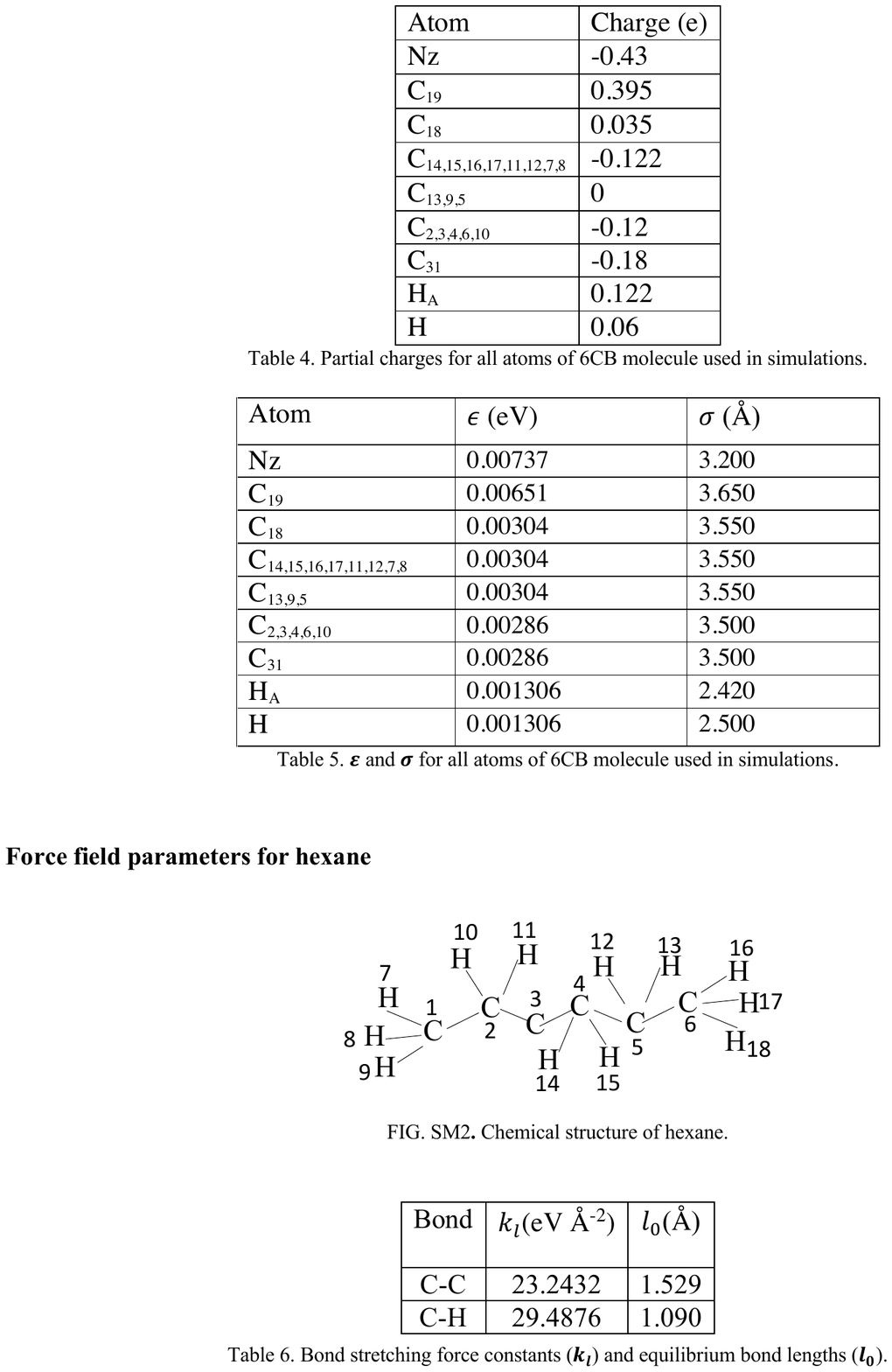}
\end{figure}
\begin{figure}[h]
    \centering
    \includegraphics[width=18 cm, height=26 cm]{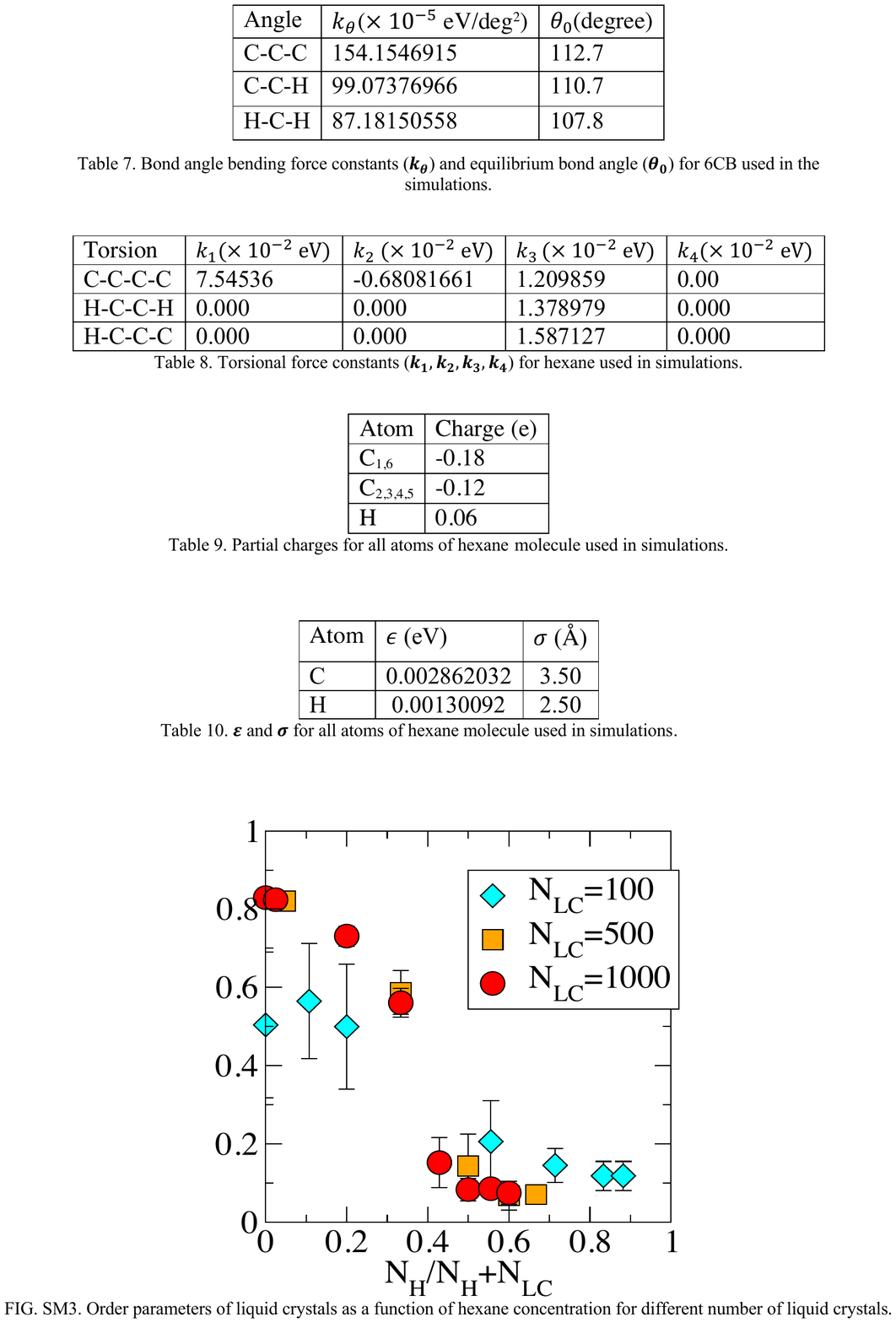}
\end{figure}
\begin{figure}[h]
    \centering
    \includegraphics[width=20 cm, height=28 cm]{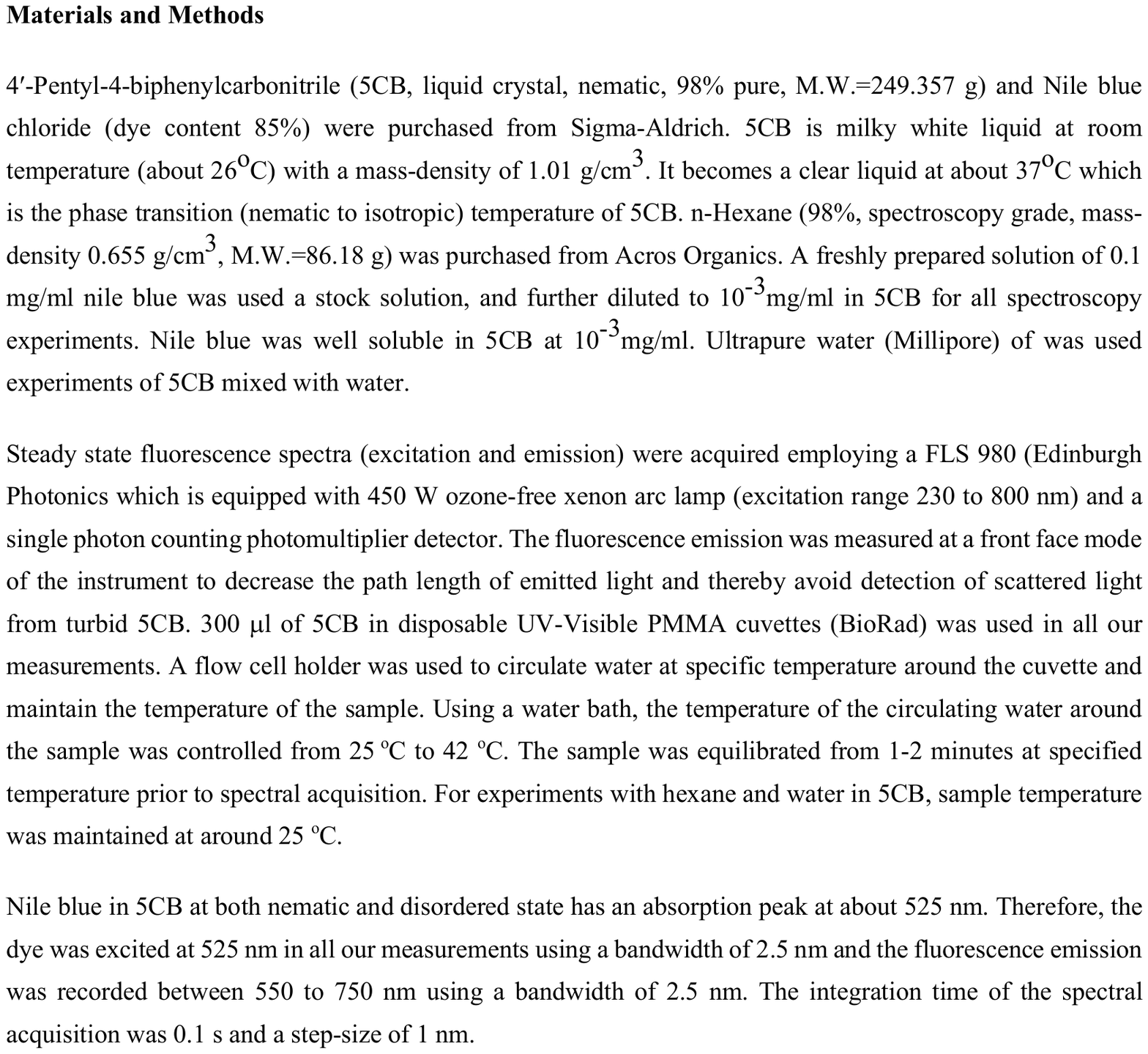}
\end{figure}

\end{document}